\documentclass[11pt]{article}
\usepackage{moriond,epsfig}
\usepackage{amssymb,mathrsfs}

\def\be{\begin{equation}}
\def\ee{\end{equation}}
\def\bc{\begin{center}}
\def\ec{\end{center}}

\newcommand{\gagg}{g_{a\gamma}}

\begin{document}

\vspace*{4cm}  
\title{SEARCH FOR SOLAR AXIONS: THE CAST EXPERIMENT AT CERN}


\author{BERTA BELTR\'AN$^6$\\
        \textit{for the CAST collaboration\\[2mm]}}

\newcommand{\CERN}{European Organization for Nuclear Research (CERN), Gen\`eve, Switzerland}
\newcommand{\Saclay}{DAPNIA, Centre d'\'Etudes Nucl\'eaires de Saclay (CEA-Saclay), Gif-sur-Yvette, France}
\newcommand{\SCarolina}{Department of Physics and Astronomy, University of South Carolina, Columbia, SC, USA}
\newcommand{\Darmstadt}{GSI-Darmstadt and Institut f\"{u}r Kernphysik, TU Darmstadt, Darmstadt, Germany}
\newcommand{\Darmstadtbis}{Technische Universit\"{a}t Darmstadt, IKP, Darmstadt, Germany}
\newcommand{\MPE}{Max-Planck-Institut f\"{u}r Extraterrestrische Physik, Garching, Germany}
\newcommand{\Zaragoza}{Instituto de F\'{\i}sica Nuclear y Altas Energ\'{\i}as, Universidad de Zaragoza, Zaragoza, Spain }
\newcommand{\Chicago}{Enrico Fermi Institute and KICP, University of Chicago, Chicago, IL, USA}
\newcommand{\Thessaloniki}{Aristotle University of Thessaloniki, Thessaloniki, Greece}
\newcommand{\Athens}{National Center for Scientific Research ``Demokritos'', Athens, Greece}
\newcommand{\Freiburg}{Albert-Ludwigs-Universit\"{a}t Freiburg, Freiburg, Germany}
\newcommand{\INR}{Institute for Nuclear Research (INR), Russian Academy of Sciences, Moscow, Russia}
\newcommand{\Vancouver}{Department of Physics and Astronomy, University of British Columbia, Vancouver, Canada }
\newcommand{\Frankfurt}{Johann Wolfgang Goethe-Universit\"at, Institut f\"ur Angewandte Physik, Frankfurt am Main, Germany}
\newcommand{\MPI}{Max-Planck-Institut f\"{u}r Physik (Werner-Heisenberg-Institut), M\"{u}nchen, Germany}
\newcommand{\Zagreb}{Rudjer Bo\v{s}kovi\'{c} Institute, Zagreb, Croatia}
\newcommand{\Pisa}{Scuola Normale Superiore, Pisa, Italy.}
\newcommand{\Lyon}{Inst. de Physique Nucl\'eaire, Lyon, France.}
\newcommand{\BNL}{Brookhaven Nat. Lab., NY-USA.}

\author{S.~ANDRIAMONJE$^2$, V.~ARSOV$^{13}$, S.~AUNE$^2$, D.~AUTIERO$^{17}$, F.~AVIGNONE$^3$, K.~BARTH$^1$, A.~BELOV$^{11}$, 
B.~BELTR\'AN$^{6}$, H.~BR\"AUNINGER$^5$, J.~M.~CARMONA$^6$, S.~CEBRI\'AN$^6$, E.~CHESI$^1$, J.~I.~COLLAR$^7$, R.~CRESWICK$^3$, 
T.~DAFNI$^4$, M.~DAVENPORT$^1$, L.~DI~LELLA$^{16}$, C.~ELEFTHERIADIS$^5$, J.~ENGLHAUSER$^5$, 
G.~FANOURAKIS$^9$, H.~FARACH$^3$, E.~FERRER$^2$, H.~FISCHER$^{10}$, J.~FRANZ$^{10}$, P.~FRIEDRICH$^5$, T.~GERALIS$^9$, 
I.~GIOMATARIS$^2$, S.~GNINENKO$^{11}$, N.~GOLOUBEV$^{11}$, R. HARTMANN$^{5}$, M.~D.~HASINOFF$^{12}$, F.~H.~HEINSIUS$^{10}$, 
D.~H.~H.~HOFFMANN$^{4}$, 
I.~G.~IRASTORZA$^{2}$, J.~JACOBY$^{13}$, 
D.~KANG$^{10}$, K.~K\"ONIGSMANN$^{10}$, R.~KOTTHAUS$^{14}$, M.~KR\v{C}MAR$^{15}$, K.~KOUSOURIS$^{9}$, M.~KUSTER$^{5,19}$, 
B.~LAKI\'C$^{15}$,
 C.~LASSEUR$^{1}$, A.~LIOLIOS$^{8}$, A.~LJUBI\v{C}I\'C$^{15}$, G.~LUTZ$^{14}$, G.~LUZ\'ON$^{6}$, D.~W.~MILLER$^{7}$, 
A.~MORALES$^{6,}$
\footnote{Deceased}, J.~MORALES$^{6}$, M.~MUTTERER$^{4}$, A.~NIKOLAIDIS$^{8}$, A.~ORTIZ$^{6}$, 
T.~PAPAEVANGELOU$^{1}$, 
A.~PLACCI$^{1}$, G.~RAFFELT$^{14}$, J.~RUZ$^{6}$, H.~RIEGE$^{4}$, M.~L.~SARSA$^{6}$, I.~SAVVIDIS$^{8}$, W.~SERBER$^{14}$, 
P.~SERPICO$^{14}$, Y.~SEMERTZIDIS$^{18}$, L.~STEWART$^{1}$, J.~D.~VIEIRA$^{7}$, J.~VILLAR$^{6}$, L.~WALCKIERS$^{1}$, 
K.~ZACHARIADOU$^{9}$ and K.~ZIOUTAS$^{8}$\\
(CAST COLLABORATION)}

\address{$^{1}$\CERN\\
         $^{2}$\Saclay\\
         $^{3}$\SCarolina\\
         $^{4}$\Darmstadt\\
         $^{5}$\MPE\\
         $^{6}$\Zaragoza\\
         $^{7}$\Chicago\\
         $^{8}$\Thessaloniki\\
         $^{9}$\Athens\\
         $^{10}$\Freiburg\\
         $^{11}$\INR\\
         $^{12}$\Vancouver\\
         $^{13}$\Frankfurt\\
         $^{14}$\MPI\\
         $^{15}$\Zagreb\\
         $^{16}$\Pisa\\
         $^{17}$\Lyon\\
         $^{18}$\BNL\\
         $^{19}$\Darmstadtbis\\
}

\maketitle


\abstracts{
Hypothetical axion-like particles with a two-photon interaction would be produced in the sun by the Primakoff process.
In a laboratory magnetic field they would be transformed into X-rays with energies of a few keV. The CAST experiment at CERN 
is using a decommissioned LHC magnet as an axion helioscope in order to search for these axion-like particles.
The analysis of the 2003 data~\cite{Andriamonje:2004hi} has shown no signal above the background, thus implying an upper 
limit to the axion-photon coupling of 
$\gagg < 1.16 \times 10^{-10}~{\rm GeV}^{-1}$ at 95\% CL for $m_{a} \lesssim 0.02~{\rm eV}$. The stable operation of the experiment
during 2004 data taking allow us to anticipate that this value will be improved.
At the end of 2005 we expect to start with the so-called second phase of CAST, when the magnet pipes will be filled with a buffer 
gas so that the axion-photon coherence will be extended. In this way we will be able to search for axions with masses up to 1 eV.
}

\section{Introduction}

 QCD is the universally accepted theory for describing the strong interactions, but it has one serious blemish: 
the so-called ``strong CP problem''. In the following we will give a brief review of it, a more general introduction on the 
subject can be found in~\cite{Jarlskog:1989bm,Turner:1989vc}.

Because of the existence of non-trivial vacuum gauge configurations, QCD has a very rich vacuum structure. All these
 degenerate
vacuum configurations of the theory are characterized by the topological winding number \textit{n} associated with them
\begin{equation}
 n=\frac{ig^3}{24\pi^2}\int d^3x\,Tr\,\varepsilon_{ijk}\,A^i(x)A^j(x)A^k(x)  
\end{equation}
where \textit{g} is the gauge coupling, $A^i$ is the gauge field, and the temporal gauge ($A^0=0$) has been used. Then, the correct 
vacuum 
state of the theory is a superposition of all these degenerate states $|n\rangle$,
\begin{equation}
|\Theta\rangle=\sum_{n}exp(-in\Theta)|n\rangle
\end{equation}
where, a priori, the angle $\Theta$ is an arbitrary parameter of the theory. States of different $\Theta$ are the physically distinct
vacua for the theory, each with a distinct world of physics built upon it. By  appropriate means the effects of this $\Theta$-vacuum can 
be recast into a single, additional non-perturbative term in the QCD Lagrangian:
\begin{equation}
\mathscr{L}_{QCD}=\mathscr{L}_{pert}+\bar{\Theta}\frac{g^2}{32\pi^2}G^{a\mu\nu}\widetilde{G}_{a\mu\nu},\;   
\bar{\Theta}=\Theta+Arg det\mathscr{M}
\end{equation}
where $G^{a\mu\nu}$ is the field strength tensor, $\widetilde{G}_{a\mu\nu}$  is its dual, and $\mathscr{M}$ is the quark mass matrix. 
This extra 
term in the QCD Lagrangian arises due to two separate and independent effects: the $\Theta$ structure of the pure QCD vacuum, and 
electroweak effects involving the quark masses. 

However, such a term in the QCD Lagrangian clearly violates CP, T and P in the case of $\bar{\Theta}\ne 0$, yet Nature has never 
exhibited
this in any experiment. Moreover, the value of the neutron electric dipole moment depends on $\bar{\Theta}$, and
the present experimental bound~\cite{Harris:1999jx} $d_{N}< 6.3\times10^{-26} {\rm e.cm}$ constrains $\bar{\Theta}$ to be less 
than (or 
of the order of) $10^{-10}$. The mystery of why the \textit{arbitrary} parameter $\bar{\Theta}$ must be so small is the strong CP 
problem.

Various theoretical attempts to solve this strong CP problem have been postulated~\cite{Jarlskog:1989bm,Dine:2000cj}, 
being the most elegant 
solution the one proposed by Peccei and Quinn in 1977~\cite{Peccei:1977ur,Peccei:1977hh}. Their idea was to make $\bar{\Theta}$ a 
dynamical variable with a classical potential that is minimized by $\bar{\Theta}=0$. This is accomplished by introducing an additional 
global, chiral symmetry, known as PQ (Peccei-Quinn) symmetry $U(1)_{PQ}$, which is spontaneously broken at a scale 
$\mathit{f}_{PQ}$. 
Immediately and independently, Weinberg~\cite{Weinberg:1977ma} and Wilczek~\cite{Wilczek:1977pj} 
realized that, because $U(1)_{PQ}$ is spontaneously broken, there should be a pseudo-Goldstone boson, ``the axion''
(or as Weinberg originally referred to it, ``the higglet''). Because  $U(1)_{PQ}$ suffers from a chiral anomaly, the axion acquires a 
small mass of the order of $m_{a}\approx~ 6~\mu {\rm eV}\left(10^{12}\,{\rm GeV}/\mathit{f}_{PQ}\right)$.

A priori the mass of the axion (or equivalently the $\mathit{f}_{PQ}$ scale) is arbitrary, but it can be
constraint using the data from various 
experiments, astrophysical considerations (cooling rates of stars) and cosmological arguments (overclosure of the Universe)
~\cite{Raffelt:1999tx,Eidelman:2004wy}. Nowadays it is believed to fall inside  the so-called ``axion mass 
window'': $10^{-6}\,{\rm eV}<~m_a<~10^{-3}{\rm eV}$. The upper limit depends on the axion-nucleon 
interaction 
that it is constrained in two different ways by the observed neutrino signal of supernova (SN)1987A~\cite{Ellis:1987pk,Raffelt:1987yt}. 
However, 
these values rely on the
model-dependent axion-nucleon coupling, they involve large statistical and systematical uncertainties, and perhaps unrecognized 
loop-holes. 
Therefore, it is prudent to consider other experimental or astrophysical methods to constraint axions in this range of parameters.

The interaction strength of axions with ordinary matter (photons, electrons and hadrons) scales~\cite{Turner:1989vc} as
$1/\mathit{f}_{PQ}$ and so the larger this number, the more weakly the axion couples. The present constraints on
its mass make the axion a weakly interacting particle, therefore a nice candidate for the Dark Matter of the 
Universe~\cite{Eidelman:2004wy}.

One generic property of the axions is a two-photon interaction of the form:
 \begin{equation}
\mathscr{L}_{a\gamma}=-\frac{1}{4}\gagg F_{\nu\mu}\widetilde{F}^{\nu\mu}a=\gagg\mathbf{E\cdot B}\,a
\end{equation}
where $F$ is the electromagnetic field-strength tensor, $\widetilde{F}$ is its dual, and ${\bf E}$ and ${\bf B}$ the electric and 
magnetic 
fields. As a consequence axions can transform into photons in external electric or magnetic fields~\cite{Primakoff},
an effect that may lead to measurable consequences in laboratory or astrophysical observations. For example, stars could produce these 
particles by transforming thermal photons in the fluctuating electromagnetic field of the stellar 
plasma~\cite{Dicus:1978fp,Raffelt:1999tx}, or axions could contribute to
the magnetically induced vacuum birefringence, interfering with the corresponding QED effect~\cite{Maiani:1986md,Raffelt:1987im}.
The PVLAS~\cite{Gastaldi} experiment apparently observes
this effect, although an interpretation in terms of axion-like particles requires a coupling strength far larger than existing limits.

\begin{figure}[ht]
\begin{center}
\epsfig{file=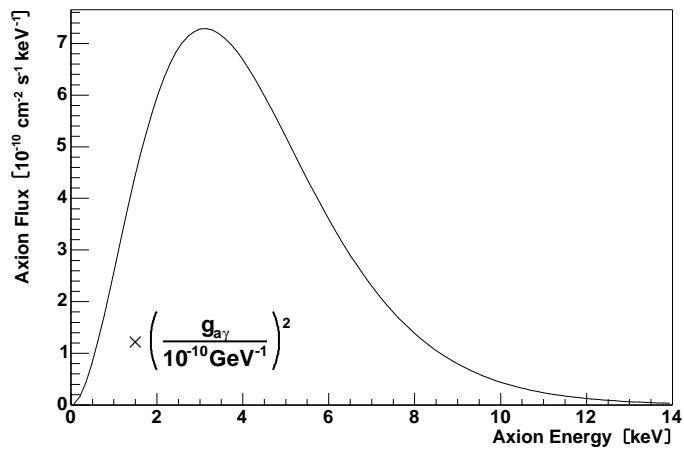,width=10cm}
\end{center}
\caption{Axion flux spectrum at the Earth}
\label{fig:axion_flux}
\end{figure}

The sun would be a strong axion source and thus offers a unique opportunity to actually detect such particles by
taking advantage of their back-conversion into X-rays in laboratory magnetic fields~\cite{Sikivie:ip}. The expected solar axion 
flux at the Earth due to the Primakoff process is:
$\Phi_a=g_{10}^2\,3.67\times10^{11}~{\rm{cm^{-2}~s^{-1}}}\;{\rm with }\ g_{10}\equiv\gagg\,10^{10}~{\rm GeV}$, 
with an approximately thermal
spectral distribution given by (Fig.~\ref{fig:axion_flux}):
\begin{equation}\label{eq:axion_flux}
\frac{d\Phi_a}{dE_a}=g_{10}^2\,3.821\times10^{10}\frac{(E_a/{\rm keV})^3}{(e^{E_a/1.103~{\rm keV}}-1)}~{\rm cm^{-2}~s^{-1}~keV^{-1}}
\end{equation}
and an average energy of 4.2~keV \footnote[1]{The spectrum in \cite{vanBibber:1988ge} has been changed to that
proposed in \cite{Creswick:1997pg}, however with a modified
normalization constant to match the total axion flux used here,
which is predicted by a more recent solar model}. The possible flux variations due to solar-model uncertainties are negligible.
Axion interactions other than the two-photon vertex would provide for additional 
production channels, but in the most interesting scenarios these channels are severely constrained, leaving the Primakoff effect as 
the dominant one \cite{Raffelt:1999tx}. In any case, it is conservative to use the Primakoff effect alone when deriving limits on 
$\gagg$.

\section{Principle of detection}\label{sec:Principle_of_detection}
A particularly intriguing application of magnetically induced axion-photon conversions is to search for solar axions using an ``axion
helioscope'' as proposed by Sikivie \cite{Sikivie:ip}. One looks at the sun through a ``magnetic telescope'' and places an X-ray
detector at the far end. Inside the magnetic field, the axion couples to a virtual photon, producing a real photon via the Primakoff 
effect: $a+\gamma_{virtual}~\rightleftharpoons\gamma$. The energy of this photon is then equal to the 
axion's total energy. The expected number of these photons that reach the X-ray detector is:
\begin{equation}
N_{\gamma}=\int\frac{d\Phi_a}{dE_a}P_{a\to\gamma}\,S\,T\,dE_a
\end{equation}
where $d\Phi_a/dE_a$ is the axion flux at the Earth as given by eq.(\ref{eq:axion_flux}), $S$ is the magnet bore 
area (${\rm cm^2}$), $T$ is the 
measurement time (s) and $P_{a\to\gamma}$ is the conversion probability of an axion into a photon. If we take some realistic numbers 
($\gagg=10^{-10}\ {\rm GeV^{-1}}, T=100\ {\rm h}$ and $S=15\ {\rm cm^2}$) this number of photons would be nearly 30 events.

The conversion probability in vacuum is given by:
\begin{equation}
P_{a\to\gamma}=\left(\frac{B\gagg}{2}\right)^2\,2L^2\,\frac{1-\cos(qL)}{(qL)^2}
\end{equation}
where $B$ and $L$ are the magnetic field and its length (given in natural units), and $q=m^2_a/2E$ is the longitudinal momentum 
difference between the axion and an X-ray of energy $E$. The conversion process is coherent when the axion and the photon fields 
remain in phase over the length of the magnetic field region.  The coherence condition states that~\cite{Lazarus:1992ry,Zioutas:1998cc} 
$qL=<\pi$ so that a coherence length of 10 m in vacuum requires $m_a\lesssim 0.02\ {\rm eV}$ for a photon energy $4.2\,{\rm keV}$. 
Coherence can be restored for a solar axion 
rest mass up to $\sim 1\,{\rm eV}$ by filling the magnetic conversion region with a buffer gas~\cite{vanBibber:1988ge} so that the 
photons
inside the magnet pipe acquire an effective mass whose wavelength can match that of the axion. For an appropriate gas 
pressure, coherence will be preserved for a narrow axion mass window.
Thus, with the proper pressure settings it is possible to scan for higher axion masses.

The first implementation of the axion helioscope concept was performed at BNL~\cite{Lazarus:1992ry}. More recently, the Tokyo axion
helioscope~\cite{Moriyama:1998kd} with $L= 2.3~{\rm m}$ and $B=3.9~{\rm T}$ has provided the limit $g_{10}<6.0$ at 95\%~CL for
$m_a\lesssim 0.03~{\rm eV}$ (vacuum) and $g_{10}<6.8$--10.9 for $m_a\lesssim 0.3~{\rm eV}$ (using a variable-pressure buffer
gas)~\cite{Inoue:2002qy}. Limits from crystal
detectors~\cite{Avignone:1997th,Morales:2001we,Bernabei:ny} are
much less restrictive.

\section{CAST experiment}

In order to detect solar axions or to improve the existing limits on $\gagg$ an axion helioscope
has been built at CERN by refurbishing a decommissioned LHC test magnet \cite{Zioutas:1998cc} which produces a magnetic field of
$B=9.0~\rm T$ in the interior of two parallel pipes of length $L=9.26~\rm m$ and a cross--sectional area $S=2\times 14.5$
cm$^2$. The aperture of each of the bores fully covers the potentially axion-emitting solar core
($\sim1/10$th of the solar radius). The magnet is mounted on a platform with $\pm 8 ^\circ$ vertical movement, allowing for
observation of the sun for 1.5 h at both sunrise and sunset. The horizontal range of  $\pm 40 ^\circ$ encompasses nearly the full
azimuthal movement of the sun throughout the year. The time the sun is not reachable is devoted to background
measurements.
A full cryogenic station is used to cool the superconducting magnet down to 1.8 K needed for its superconducting operation
\cite{Barth:2004cx}. The hardware and software of the tracking system have been precisely calibrated, by means of geometric
survey measurements, in order to orient the magnet to any given
celestial coordinates. The overall CAST pointing precision is better~\cite{filming} than 0.01$^\circ$  including all
sources of inaccuracy such as astronomical calculations, as well
as spatial position measurements.
At both ends of the magnet, three different detectors have searched for excess X-rays from axion conversion in the magnet
when it was pointing to the sun. Covering both bores of one of the magnet's ends, a conventional Time Projection Chamber (TPC) is
looking for X-rays from ``sunset'' axions. At the other end, facing ``sunrise'' axions, a second smaller
gaseous chamber with novel MICROMEGAS (micromesh gaseous structure -- MM) \cite{Giomataris:1995fq} readout is placed behind
one of the magnet bores, while in the other one, a X-ray mirror telescope is used with a Charge Coupled Device~\cite{Struder:2001bh}
 (pn-CCD) as
the focal plane detector. Both the pn-CCD and the X-ray telescope are prototypes developed for X-ray astronomy \cite{abrixas}.
The X-ray mirror telescope can produce an ``axion image'' of the sun by focusing the photons from axion conversion to a
$\sim6\,{\rm mm}^2$ spot on the pn-CCD. The enhanced signal-to-background ratio substantially improves the sensitivity
of the experiment.
\subsection{First phase of CAST}
During the years 2003 and 2004 the CAST experiment has gone through the so-called first phase, where the data has been taken with 
vacuum inside the magnetic field area, so that we were sensitive to 
axion masses up to $m_a\lesssim 0.02\,{\rm eV}$ as explained in section~\ref{sec:Principle_of_detection}. 

\subsection{Second phase of CAST}
In order to extend the range of axion masses to which we are sensitive, the 
magnet pipes will be filled with Helium gas in phase II. As explained in section~\ref{sec:Principle_of_detection}, 
a gas with a given pressure will provide a refractive 
photon mass so that the coherence of the photon and axion fields will be restored for a certain range of axion masses.
The second phase of the experiment is very challenging because, for the first time, a laboratory experiment will search for axions 
in the
theoretically motivated range of axion parameters (see Fig.~\ref{exclusion}).

Data taking for this second phase it is scheduled to begin at the end of 2005, with low  pressure ${\rm^{4}He}$ gas inside the 
pipes at 1.8 K, the magnet's operating temperature.  There is a limit in the pressure that we can reach with ${\rm ^{4}He}$ before it 
liquefies, so in order to be able to extend the mass axion searches up to $\sim 0.82\,{\rm eV}$ we will have to switch to ${\rm ^{3}He}$,
 which
has a higher vapor pressure. These steps are scheduled to occur during 2006 and 2007.

Beyond these plans CAST could search for axions with still higher 
masses up to $\sim 1.4\,{\rm eV}$ with the actual set-up, by installing thermally 
isolated gas cells inside the magnet bores. This would allow us to work 
at higher temperatures ($\sim$ 5.4 K) so that we could reach higher 
pressures and densities of the ${\rm^{4}He}$ buffer gas.

\section{Data analysis and first results}
\subsection{2003 data tacking}
CAST operated for about 6 months from May to November in 2003, during most of which time
at least one detector was taking data.
The results~\cite{Andriamonje:2004hi} presented in this paper were obtained after the analysis of the data sets listed in 
Table \ref{datasets}.
\begin{table}[b] \centering \footnotesize
\caption{ Data sets included in our result. \label{datasets}}
\begin{tabular}{cccccccc}
\\ \hline\hline
\multicolumn{1}{c}{Data set} & \multicolumn{1}{c}{Tracking} & \multicolumn{1}{c}{Background} &
\multicolumn{1}{c}{$(g^4_{a\gamma})_{\rm best fit}$ ($\pm 1\sigma$
error)} & \multicolumn{1}{c}{$\chi^2_{\rm null}$/d.o.f} &
\multicolumn{1}{c}{$\chi^2_{\rm min}$/d.o.f} &
\multicolumn{1}{c}{$g_{a\gamma}$(95\%)} \\ & exposure(h) & exposure(h) &
$(10^{-40}$ GeV$^{-4})$ & & & $(10^{-10}$ GeV$^{-1})$\\ \hline
 TPC & 62.7 & 719.9 & $-1.1\pm3.3$ & 18.2/18 & 18.1/17 & $1.55$\\
MM set A & 43.8 & 431.4 & $-1.4\pm 4.5$ & 12.5/14 & 12.4/13 &
$1.67$\\ MM set B & 11.5 & 121.0 & $2.5\pm 8.8$ & 6.2/14 & 6.1/13
& $2.09$
\\ MM set C & 21.8 & 251.0 & $-9.4\pm 6.5$
& 12.8/14 & 10.7/13 & $1.67$
\\ pn-CCD &
121.3 & 1233.5 & $0.4 \pm 1.0$ & 28.6/20 & 28.5/19 & $1.23$\\
\hline
\end{tabular}\end{table}
An independent 
analysis was performed for each data set. Finally, the results
from all data sets are combined.

An important feature of the CAST data treatment is that the
detector backgrounds are measured with $\sim$10 times longer
exposure during the non-alignment periods.
The use of these data to estimate and subtract the true
experimental background during sun tracking data is the most
sensitive step in the CAST analysis. To assure the absence of
systematic effects, the main strategy of CAST is the use of three
independent detectors with complementary approaches. In the event
of a positive signal, it should appear consistently in each of the three detectors when it is pointing at the
sun.
In addition, an exhaustive recording of experimental parameters
was done, and a search for possible background dependencies on
these parameters was performed.
A dependence of the TPC background on the magnet position was
found, caused by its relatively large spatial movements at the far
end of the magnet, which resulted in appreciably different
environmental radioactivity levels. Within statistics, no such
effect was observed for the sunrise detectors which undergo a much
more restricted movement. To correct for this systematic effect in
the TPC data analysis, an effective background spectrum is
constructed only from the background data taken in magnet
positions where sun tracking has been performed and this is
weighted accordingly with the relative exposure of the tracking
data.
Further checks have been performed in order to exclude any
possible systematic effect. They were based on rebinning the data,
varying the fitting window, splitting the data into subsets and
verifying the null hypothesis test in energy windows or areas of
the detectors where no signal is expected. In general, the
systematic uncertainties are estimated to have an effect of less
than $\sim$10\% of the final upper limits obtained.

For a fixed $m_a$, the theoretically expected spectrum of
axion-induced photons has been calculated and multiplied by the
detector efficiency curves 
of the detectors, including all
hardware and software efficiency losses, such as window
transmissions (for TPC and MM), X-ray mirror reflectivity (for
pn-CCD), detection efficiency and dead time effects.
These spectra, which are proportional to $g_{a\gamma}^4$, are
directly used as fit functions to the experimental subtracted
spectra (tracking minus background) for the TPC and MM. For these
data, the fitting is performed by standard $\chi^2$ minimization.
Regarding the pn-CCD data, the analysis is restricted to the small
area on the pn-CCD where the axion signal is expected after the
focusing of the X-ray telescope. During the data taking period of
2003 a continuous monitoring of the pointing stability
of the X-ray telescope was not yet possible, 
therefore a signal area larger than the size of the sun spot had
to be considered. Taking into account all uncertainties of the
telescope alignment, the size of the area containing the signal
was conservatively estimated to be $34\times71\,$pixels
($54.3\,{\rm mm}^2$). As in the other detectors, the background
is defined by the data taken from the same area during the
non-tracking periods, but, in addition, the background in the
signal area was also determined by extrapolating the background
measured during tracking periods in the part of the pn-CCD not
containing the sun spot. Both methods of background selection led
to the same final upper limit on the coupling constant
$g_{a\gamma}$. The resulting low counting statistics in the pn-CCD
required the use of a likelihood function in the minimization
procedure, rather than a $\chi^2$-analysis.
The best fit values of $g_{a\gamma}^4$ obtained for each of the
data sets are shown in Table \ref{datasets}, together with their
1$\sigma$ error and the corresponding $\chi^2_{\rm min}$ values
and degrees of freedom. 
\begin{figure}[t]
\begin{center}
\psfig{figure=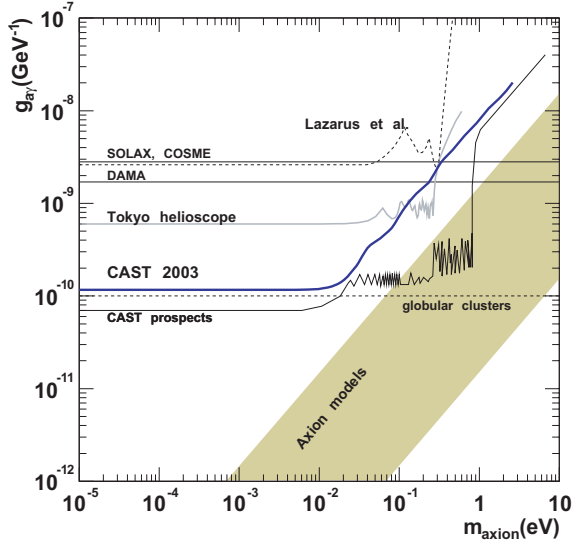,width=75mm}
 \caption{Exclusion limit (95\% CL) from the CAST 2003 data compared with other
 constraints dicused in section 2. The shaded band represents
 typical theoretical models. Also shown is the future CAST
 sensitivity as foreseen in the
 experiment proposal.
 \label{exclusion}}
 \end{center}
\end{figure}
Each of the data sets is individually compatible with the absence
of any signal as can be seen from the $\chi^2_{\rm null}$ values
shown in Table \ref{datasets}.
The excluded value of $g_{a\gamma}^4$ was conservatively
calculated by taking the limit encompassing 95\% of the physically
allowed part (i.e. positive signals) of the Bayesian probability
distribution with a flat prior in $g_{a\gamma}^4$. The described
procedures were done using $g_{a\gamma}^4$ instead of
$g_{a\gamma}$ as the minimization and integration parameter
because the signal strength (i.e. number of counts) is
proportional to $g_{a\gamma}^4$.
The 95\% CL limits on $g_{a\gamma}$ for each of the data sets are
shown in the last column of Table \ref{datasets}. They can be
statistically combined by multiplying the Bayesian probability
functions and repeating the previous process to find the combined
result for the 2003 CAST data:

\begin{equation}
 g_{a\gamma} < 1.16\times 10^{-10}{\rm GeV}^{-1} (95\%{\rm CL}).
\end{equation}

Thus far, our analysis was limited to the mass range
$m_a\lesssim0.02$~eV where the expected signal is mass-independent
because the axion-photon oscillation length far exceeds the length
of the magnet. For higher $m_a$ the overall signal strength
diminishes rapidly and the spectral shape differs. Our procedure
was repeated for different values of $m_a$ to obtain the entire
95\% CL exclusion line shown in Fig.~\ref{exclusion}.

\subsection{2004 data tacking}
The data taken from 2004 have not yet been fully analyzed. However, the stable operation of the experiment allowed the CAST collaboration
to take enough high-quality data to anticipate that the final sensitivity will be close to the value presented in the CAST proposal (see 
Fig.~\ref{exclusion})

\section{Summary}
The origin of the axion as a particle that solves the strong CP problem has been reviewed. Some properties of this pseudoescalar 
particle have been pointed out, among them the fact that it can transform into a photon in external electric or magnetic fields, 
this being
the only property of the axion on which CAST relies. The CAST experiment and its first results~\cite{Andriamonje:2004hi}have been 
presented.
Our limit improves the best previous laboratory
constraints \cite{Moriyama:1998kd} on $g_{a\gamma}$ by a factor 5
in our coherence region $m_a\lesssim0.02$~eV. 
A higher sensitivity is expected from the 2004 data
with improved conditions in all detectors, which should allow us
to surpass the astrophysical limit.
%
In addition, starting in 2005, CAST plans to take data with a
varying-pressure buffer gas in the magnet pipes, in order to
restore coherence for axion masses above 0.02 eV. The extended
sensitivity to higher axion masses will allow us to enter into the
region shown in Fig.~\ref{exclusion} which is especially motivated
by axion models \cite{Kaplan:1985dv}.

\end{document}